# Tuning the Mechanical Properties in Model Nanocomposites: Influence of the Polymer-Filler Interfacial Interactions

*Chloé Chevigny [a,†], Nicolas Jouault [a,‡*], Florent Dalmas [b], François Boué [a] and Jacques Jestin [a]*

[a] Laboratoire Léon Brillouin, CEA Saclay 91191 Gif-sur-Yvette Cedex France

[b] Institut de Chimie et des Matériaux Paris-Est, UMR 7182 CNRS/Université Paris-Est Créteil, 2-8 rue Henri Dunant 94320 Thiais France

[†] Present address: Stranski-Laboratorium für Physikalische und Theoretische Chemie, Institut für Chemie, Technische Universität Berlin, Sekr. TC 9, Str. des 17. Juni 124, 10623 Berlin, Germany

[‡] Present address: Laboratoire Matière et Systèmes Complexes (MSC), UMR 7057 CNRS-Université Paris 7-Diderot, Bâtiment Condorcet - CC 7056, 75205 Paris Cedex 13, France

[*] Corresponding author

nicolas.jouault@univ-paris-diderot.fr

KEYWORDS: nanocomposites, SANS, mechanical reinforcement, interfacial interaction

**ABSTRACT**

This paper presents a study of the polymer-filler interfacial effects on filler dispersion and mechanical reinforcement in Polystyrene (PS) / silica nanocomposites by direct comparison of two model systems: un-grafted and PS-grafted silica dispersed in PS matrix. The structure of nanoparticles has been investigated by combining Small Angle Neutron Scattering (SANS) measurements and Transmission Electronic Microscopic (TEM) images. The mechanical properties were studied over a wide range of deformation by plate/plate rheology and uni-axial stretching. At low silica volume fraction, the particles arrange, for both systems, in small finite size non-connected aggregates and the materials exhibit a solid-like behavior independent of the local polymer/fillers interactions suggesting that reinforcement is dominated by additional long range effects. At high silica volume fraction, a continuous connected network is created leading to a fast increase of reinforcement whose amplitude is then directly dependent on the strength of the local particle/particle interactions and lower with grafting likely due to deformation of grafted polymer.

**Introduction**

Reinforcement of amorphous polymers and elastomers using inorganic particles is still an open challenge, from both industrial and fundamental points of view. According to the characteristics of the filler (nature, shape, concentration) and of the matrix (interaction potential with the filler), many systems with many associated mechanical properties can be found [1]. The enhancement of the composite properties can be induced by increasing the specific contact area between the filler and the matrix, i.e., by reducing the size of the particle filler. This reduction can be made by using nano-sized particles, but this would produce an inconvenient extent of clustering. Controlled clustering by specific filler/matrix interaction enables the control of the hierarchical organization of fillers in the matrix and opens the way to new mechanical properties. Together with advances in characterization methods, this leads to the creation of advanced composites with nano-fillers, from classical carbon black [2] to silica [3] and up to carbon nano-tubes [4]. Most of studies use rheological approaches of various mechanical solicitations, and show various kinds of rheological transitions related to reinforcement. Up to date, two main contributions to reinforcement stand out, borrowing both from earlier research and from more recent theory. The first contribution is linked to the filler network structure: the idea is that the particles can form a connected network [5-6] as in the percolation model. The second contribution is attributed to the interfacial interaction between the polymer chains (matrix) and the particles (fillers): particles may involve a new elastic contribution and slowing-downs of the chains mobility, leading to a gradient of glass-transition temperature $T_g$ from filler surface to polymer bulk and going as far as glasslike zones around the particles, increasing the rigid phase volume fraction [7-8]. Modifications of the polymer chains conformation are also proposed to contribute the elastic modulus by bridging effects of chains between the fillers [9-10-11]. The experimental distinction between these two classes of contributions, filler network and interfacial interactions, is difficult because they are often correlated to each other. The native morphology of complex fillers (such as carbon black or fumed silica) could affects the buildup, the geometry, the connectivity of the filler network, and the related interfacial interactions. The interaction potential between the chain and the filler is also associated with the filler form factor and can thus induce complex pictures, even when starting from simple individual nanoparticles.

In order to control the clusters formation or the particles dispersion, the interfacial interaction between particles and polymer could be tuned via an external trigger like magnetic field [12], or via an internal one, modifying the particles surface: grafting surfactants [13] or longer molecules, up to polymer chains via the "grafting from" or "grafting to" methods [14]. Grafting polymer chains on the fillers can be used to control the internal organization in the composites and study the effect of the change in interfacial interactions on particles dispersion and mechanical properties of the nanocomposites. Recent results show that mechanical reinforcement is mainly based on the nanoparticles [15, 16, 17, 18] (no matrix contribution), especially for individual particles dispersion in which solid-like behavior can be related to the direct contact between grafted brushes [19-20]. However, a general rule describing the mechanical behavior as a function of several particles dispersion, which can be individual, finite size aggregates (connected or not), or larger agglomerates, remains currently to be established. To progress in the understanding of the different contributions (filler effect, chains bridging or chains dynamics), it is essential to be able to dissociate them more finely from the local sizes (at the nanoparticle/polymer interface) to larger ones. Previous works only focused on one specific contribution. Cassagnau *et al.* [6, 21] made rheology measurements on nanocomposites filled with both un-grafted and grafted silica, but without precise knowledge of the dispersion state, which make difficult the interpretations of the origin of the observed reinforcement. Other studies [22, 23, 24] have focused on interfacial effects on mechanical properties once again without precise information about the particles dispersion.

Our interest in the present paper is to dissociate the interfacial contributions from the filler network one. We directly compare the filler structure and mechanical properties of similar silica/polystyrene (PS) nanocomposites differing only by surface particle state: in the first case, silica particles are unmodified [25], while in the second case, the silica surface is grafted with PS chains. Except for the grafting step, the silica particles are exactly similar (Nissan-ST in Dimethylacetamide) and the nanocomposites processing (evaporation time, temperature...) is kept unchanged. We first present the characterization of the grafting steps using scattering experiments (Small Angles X-rays and Neutrons Scattering, SAXS and SANS) and neutron contrast variation method. Second, the (grafted and un-grafted) particles dispersion inside the polymeric matrix is determined at local scale with SANS, as a function of the silica particles concentration. Dispersion at larger scales (up to the micron) is investigated using imaging technique (Transmission Electronic Microscopy, TEM). Thirdly, mechanical behavior is studied both at low (oscillatory shear measurements ARES) and high deformations (stretching experiments) on both systems, which enables us to directly compare them and finally discuss the correlation between the particle dispersion and the mechanical behavior mediated by two situations of local polymer-filler interaction.

**Materials and Methods**

1. *Synthesis of polystyrene (PS)/grafted particles*

We use a colloidal silica sol of Nissan-ST particles, kindly provided by Nissan Chemicals, dispersed in the organic solvent DiMethylAcetamide (DMAc). To graft PS chains on the particles surface, we use a "grafting from" method associated with a nitroxide-mediated polymerization (NMP) [26]. First, the initiator is grafted on the particles, and then polymerization is started, in a controlled way, from the surface. Details of the initiator-grafting procedure and polymerization were described in previous papers [27] for Ludox silica particles. The process here is exactly the same: the sol is silanized using AminoPropylTriEthoxySilane, and the SG1-based initiator is then grafted over. After each grafting step the solution is purified of the un-reacted molecules by ultrafiltration. The polymerization takes place for 4 hours at 120°C, under nitrogen flow. The control of the polymerization is guaranteed via the addition of free initiator (MAMA-SG1, provided by Arkema) to the reaction mixture. All the parameters of the reaction (experimental conditions, silica concentration and monomer quantity) were optimized to keep the colloidal stability of the sol while having the best possible conversion. The success of the grafting process was checked by Thermo-gravimetric Analysis (TGA), by measuring the mass of grafted polymer. After the polymerization, the grafted particles are separated from un-tethered chains by ultra-filtration. These chains, characterized by Size Exclusion Chromatography (SEC), have a molar mass $M_n$ of 29 000 g/mol, which can reasonably be assimilated to the mass of the grafted chains [27, 28]. After the polymerization process, the control of the sol stability and the structural parameters of the grafted objects have been checked by SANS using contrast-matching technique (described below).

2. *Nanocomposites preparation*

The preparation is the same as described in the previous paper [25] for the native (un-grafted) particles. The matrix polymer (Polystyrene, $M_w$=280 000 g/mol, PDI = 2, purchased from Sigma-Aldrich) is dissolved at 10%v/v in DMAc, and mixed with silica solutions at different concentrations (from 0 to 30%v/v). Concerning the grafted particles, due to the synthesis constraints, the concentration of silica sol is fixed at 1%v/v. The volume is varied to change the quantity of silica added. The mixtures are stirred (using a magnetic rod) for 2 h. They are then poured into Teflon molds (5 cm x 5 cm x 2.5 cm) and let cast in an oven at a constant temperature of 130 °C during 8 days. This yields dry films of dimensions of 5 cm x 5 cm x 0.1 cm (i.e., a volume of 2.5 cm$^3$). We should note that, due to the limited volume of the mold and the limited concentration of grafted silica solution, the maximal silica volume fraction reached is much lower for the grafted particles (about 12%v/v) than for the native particles (about 30%v/v).

3. *Small Angle Scattering*

Small Angle Neutron Scattering (SANS) measurements were performed at the Laboratoire Léon Brillouin (LLB, CEA Saclay, France) on the SANS spectrometer PAXY. Three configurations were used: the first one with wavelength 15 Å, sample-to-detector distance of 6.70 m, and a collimation distance of 5.00 m, and the second with wavelength 6 Å, sample-to-detector distance of 6.70 m, and a collimation distance of 2.50 m and the last one with wavelength 6 Å, sample-to-detector distance of 3.00 m, and a collimation distance of 2.50 m, corresponding to a total Q-range of $2.10^{-3}$ Å$^{-1}$ to 0.1 Å$^{-1}$. The wavelength distribution Δλ/λ is around 10%. Data processing was performed with a homemade program following standard procedures with H$_2$O as calibration standard. Small deviations, found in the spectra at the overlap of two configurations, are due to different resolution conditions and (slight) remaining contributions of inelastic, incoherent, and multiple scattering. To get the cross-section per volume in absolute units (cm$^{-1}$), the incoherent scattering cross-section of H$_2$O was used as a calibration. It was estimated from a measurement of the attenuator strength, and of the direct beam with the same attenuator. The incoherent scattering background, mainly due to protons of the polystyrene matrix, was subtracted using a blank sample with no silica particles.

For the silica solution, Small Angle X-rays Scattering (SAXS) measurements were performed on SWING beam-line (Soleil synchrotron) and were recorded using 2D AVIEX CDD camera at energy of 7 keV with 2 samples-to-detector distances: 6.5m and 1.8m, leading to a Q-range from $1.8.10^{-3}$ Å$^{-1}$ to 0.15 Å$^{-1}$.

4. *Transmission Electronic Microscopy*

The samples were cut at room temperature by ultra-microtomy using a Leica Ultracup UCT microtome with a diamond knife. The cutting speed was set to 0.2mm.s$^{-1}$. The thin sections of about 40 nm thickness were floated on deionized water and collected on a 400-mesh copper grid. Transmission electron microscopy was performed on a Philips Tecnai F20 ST microscope (field-emission gun operated at 3.8 kV extraction voltage) operating at 200 kV. Precise scans of various regions of the sample were systematically done, starting at a small magnification which was then gradually increased.

5. *Low deformation: plate/plate rheometer (ARES)*

Shear tests, corresponding to low deformation levels (0.5%), were carried out in the dynamic mode in strain-controlled conditions with a plate-plate cell of an ARES spectrometer (Rheometrics-TA) equipped with an air-pulsed oven. This thermal environment ensures a temperature control within 0.1 °C. The samples were placed between the two plate (diameter 1mm) fixtures at high temperature (160 °C), far above the glass transition, put under slight normal stress (around 0.5 N), and temperature was decreased progressively, while gently reducing the gap to maintain a constant low normal stress under thermal retraction. The zero gap is set by contact, the error on sample thicknesses is thus minimal and estimated at ±0.010mm with respect to the indicated value. The shear amplitude was fixed to 0.5% to stay below the limit of linear deformation. Samples are temperature-stabilized for 30 minutes before starting measurements. In dynamic mode, the frequency range is 0.5-100 rad/s for different temperatures (from 190°C to 120 °C), and time-temperature superposition is applied. The obtained multiplicative factor can be adjusted to WLF law [29] as follows:

$$log(a_T) = \frac{C_1 \cdot (T_{ref} - T)}{C_2 + T - T_{ref}}$$

where $a_T$ is the multiplicative factor, $T_{ref}$ is the reference temperature of the master curve (in our case 143 °C), T the temperature of the measurement, $C_1$ and $C_2$ the WLF parameters. We find $C_1$=6.72 and $C_2$=98.03 °C for both systems, which is commonly obtained for PS samples.

6. *Uni-axial stretching*

For uni-axial stretching, rectangular films are cut 2 cm x 0.5 cm x 0.1 cm and they are carefully sanded down to a constant thickness. Samples are stretched up at $T_g+20°C$ in an oil bath to a predefined deformation with a constant rate deformation of $0.005s^{-1}$ using a homemade apparatus. Glass-transition temperatures $T_g$s were determined by Differential Scanning Calorimetry DSC and gives for un-grafted systems $\Delta T_g = +1°C$ at 5% v/v, $\Delta T_g = +4°C$ at 15% v/v and $\Delta T_g = 0°C$ for the grafted system. Only a small change is observed as a function of filler incorporation as already observed in [30]. The tensile force is measured with a HBM Q11 force transducer and converted to real stress by dividing the force by the assumed cross section during the elongation. The elongation ratio is defined as $\lambda=L/L_0$.

**Results**

1. *Native silica particles*

The native silica particles (fillers of both systems) have been first characterized dispersed in solvent (DMAc) by SAXS. The scattering curve of a diluted solution 0.1% v/v is presented in Figure 1. These particles have theoretically a mean radius equal to 5 nm (value given by Nissan Chemical) but the calculated sphere form factor corresponding to this size cannot fit correctly the scattering curve: the apparent radius appears to be larger, indicating that particles are slightly aggregated. Meanwhile, we observe a maximum in the low-Q region corresponding to the mean distance (D~$2\pi/Q^*$~110 nm) between finite-size objects. Assuming that the objects are included in a cubic network of mesh size D, we can extract an estimation of the number of primary particles per object $N_{agg}$ with:

$$N_{agg} = D^3 \Phi_{SiO2}/(4/3\pi R^3) \qquad (1)$$

Where $\Phi_{SiO2}$ is the volume fraction of particles, and R is the radius of the primary particle, equal to 5 nm. Such calculation gives an aggregation number of 3-4: in solution, primary particles arrange themselves into small finite-size clusters of 3-4 units. These clusters are compact and can be then fitted with a 6.0nm-radius polydisperse sphere form factor (red curve), using a log-normal distribution in size of $\sigma = 0.3$.

[Figure 1]

2. *Grafted Silica particles*

After the polymerization process, the PS-grafted particles have been purified to remove the free chains and can be then characterized in solution using SANS and contrast variation method. The contribution of the grafted polymer can be firstly suppressed with a specific mixture of deuterated and normal solvent (15/85 % v/v). The corresponding scattering curve (black circles), presented Figure 2(a), is very close to the scattering signal of the silica particles in solution (red curve) indicating that we observe the same 3-4 units clusters. In addition, we observe a maximum in the low-Q region corresponding to the structure factor between the grafted particles. The structure factor S(Q) can be calculated by dividing the total scattered intensity I(Q) by the calculated form factor (P(Q), red curve), giving a nice repulsive structure factor (shown in insert) illustrating the stabilization and the absence of additional aggregation of particles during the polymerization. To observe the grafted layer on the same sample, the contribution of the silica core can be suppressed with another mixture of deuterated and normal solvent (47/53 % v/v). The scattering curve obtained in this contrast condition normalized by the structure factor S(Q) determined above, presented in Figure 2(b), is then rather different from the previous one. In the low-Q region, we observe a plateau followed by a rapid decrease of the intensity corresponding to the finite global size of the grafted corona. As expected, the curve is shifted toward the lower Q-range in comparison to the silica scattering, meaning that grafted corona is larger than silica core. In the intermediate Q-region, we observe a smooth oscillation corresponding to the thickness of the grafted polymer layer and, at larger Q, a decrease of the intensity as $Q^{-2}$ characteristic of the local scattering contribution of the grafted chains. To model the grafted layer, we use a Gaussian-chains model described by Pedersen from which we can extract the relevant parameters characterizing the grafted objects. The details of the modeling process, as well as equations and references for the Pedersen model, are described in our previous works [27]. The best fit result of the modeling (blue curve on Figure 2(b)) gives us the following parameters: particle volume fraction of 0.75% v/v, silica core radius of 6.0 nm and polydispersity of 0.3, gyration radius of the grafted chains of 7.8 nm, and 212 chains per particle. Thanks to SANS and contrast variation technique, we could check that PS-grafted silica particles are well-dispersed in solution and can be used to prepare nanocomposites.

**[Figure 2]**

3. *Filler dispersion*

The dispersion of the silica nanoparticles in the polymeric matrix is investigated by SANS measurements and TEM observations. The two techniques are complementary and provide an extensive insight into the particles distribution over a wide range of observation scale from the nanometer (local) to the micron scale. The nanocomposites filled with grafted particles are a three-component system: the PS matrix, the silica particles and the grafted PS "hairs" on the particles. Because the same component is used for the matrix and for the hairs, there is no contrast between grafted and free chains and, in SANS, we only "see" the silica particles inside the polymeric medium. The Figure 3 presents the scattered intensities of nanocomposites filled with un-grafted (Figure 3(a)) and PS-grafted particles (Figure 3(b)) at low silica concentration $\Phi_{SiO2}$, respectively at 6.65 and 4.30 % v/v.

[Figure 3]

At high Q-values (0.03-0.1 Å$^{-1}$), the curves superimpose well and the intensity scales as Q$^{-4}$, characteristic of sharp spherical interfaces (here the interface between the silica surface and the polymer). Then, as Q decreases below 0.03 Å$^{-1}$, the scattered intensity increases (while the form factor of the primary particles reaches a plateau, see Figure 1), suggesting that objects larger than the single particles are present in the composite: aggregates are formed. At intermediate Q-range, a change in slope is observed with a more pronounced shoulder for un-grafted systems. This shoulder comes from an intra-particle structure factor corresponding to the direct contact between native particles inside aggregates. Then, the slope of this increase is characteristic of the fractal dimension $D_f$ of the aggregates, with a stronger increase of the intensity of the aggregates in the un-grafted case. But, for both cases, the scattered intensities finally reach a plateau at the lowest Q-values (until 3.10$^{-3}$ Å$^{-1}$), indicating the finite size of these fractal aggregates. The scattering intensity of such fractal aggregates present three different behaviors in three different Q ranges separated by tow cut-offs Qc1 and Qc2 [31]:

(i) at very large Q (for Q > Qc2), the scattering can be identified with the form factor P(Q,R) of the primary particles:

$$I(Q) = \Phi \Delta\rho^2 \frac{\int_0^\infty P_{particle}(Q,R) l(R,\sigma) R^3 dR}{\int_0^\infty R^3 L(R,\sigma) dR}$$

(ii) at intermediate Q (for Qc1 > Q > Qc2), the scattering intensity scales as

$$I(Q) \sim Q^{-D_f}$$

(iii) at low Q (for Q > Qc1), the scattering intensity is equal to:

$$I(Q) = \Phi \Delta\rho^2 N_{agg} P(Q \to 0, R) \qquad (2)$$

where $N_{agg}$ is the number of single particles inside an aggregate, $D_f$ is the fractal dimension of the aggregates, $P_{particle}(Q,R)$ is the form factor of the native silica particles deduced from the Figure 1 (R= 6 nm) and $L(R,\sigma)$ is the log-normal distribution of the single particles radius (variance $\sigma=0.3$). As the particles are strongly diluted in the film, interactions between aggregates are considered negligible: the resulting inter-aggregate structure factor is close to 1. The total scattered intensity I(Q) is then directly related to the aggregates form factor:

$$I(Q) \approx \phi_{SiO_2} \Delta\rho^2 V_{agg} P_{agg}(Q) \qquad (3)$$

The volume fraction $\Phi_{SiO2}$ is fixed at the experimental values (respectively 4.30 and 6.65 % v/v, determined by TGA measurements) and only the contrast term $\Delta\rho^2$, $D_f$ and $N_{agg}$ are to be fitted. The best fit is represented in black continuous line in Figure 3(a) for the un-grafted system and Figure 3(b) for the grafted one. The fitted curves agree quite well with the experimental ones. For the un-grafted case, the fitting parameters are $\Delta\rho^2=3.88\ 10^{20}$ cm$^{-4}$, $D_f = 1.75$ and $N_{agg}=9$. The value of the contrast term $\Delta\rho^2$ deduced from this calculation corresponds exactly to the difference between the theoretical scattering length density (SLD) of the silica (3.40 $10^{10}$ cm$^{-2}$) and the SLD of the polymer (1.43 $10^{10}$ cm$^{-2}$), meaning that we observe silica aggregates inside a PS matrix. For the grafted particles, we obtain a fractal dimension $D_f$ of 1.1, an aggregation number $N_{agg}$ of 7 and a contrast term $\Delta\rho^2$ of 2.46 $10^{20}$ cm$^{-4}$. The value of $\Delta\rho^2$ found in this former case implies a mean SLD of the aggregates lower (3.00 $10^{10}$ cm$^{-2}$) than the one of pure silica, indicating that they are made of both silica and PS chains (grafted on the particles). Small discrepancies between data and fit are also observed, especially for the un-grafted system, which can be related to the intra aggregate structure factor $S_{agg\ intra}(Q)$ coming from the interaction between the silica particles inside the aggregates and not taken into account in our calculations. The Figure 3(c) presents the corresponding intra aggregate structure factor $S_{agg\ intra}(Q)=I(Q)/P_{agg}(Q)$ for both systems. A deeper correlation hole is observed for the un-grafted system: aggregates are denser, leading to stronger and better-defined attractive interactions between particles which are in contact. For the grafted system, these interactions are modulated and attenuated by the polymer layer as illustrated by the absence of the correlation hole. However, in the small-Q domain, the two structure factors superimpose nicely with a flat plateau, which confirms the absence of correlation between aggregates at these concentrations.

[Figure 4]

The TEM pictures (Figure 4(a) for un-grafted system at 6.65% v/v and 4(b) for the grafted system at 4.30% v/v of silica) show the fillers dispersion on a wider scale and confirm SANS measurements: silica particles arrange in small fractal aggregates not directly connected in both cases. The effect of particles dilution is visible for the grafted system, where aggregates appear to be smaller which is in agreement with SANS results (slightly lower intensity at the lowest Q-values, indicating smaller objects). Single particles are also observed in the grafted case and could simply be attributed to a better compatibility via the polymer "hairs" while un-grafted aggregates appears to be more homogenously distributed in size. We have performed the image analysis for un-grafted system at 6.65% v/v from which we can extracted the following parameters for the aggregates : $R_g$ = 26nm, $N_{agg}$ = 23 and $D_f$ = 1.6 (see Supporting Information). If the agreement is good with the SANS determination for the fractal dimension, we observed a small discrepancy for the aggregation number which can coming form cutting depth effects."

[Figure 5]

The Figure 5(a) presents the SANS curves, at higher silica volume fractions $\Phi_{SiO2}$, for the un-grafted system at 15.4% v/v and for the grafted system at 11.6% v/v. Intensities are normalized by $\Phi\Delta\rho^2$ using values deduced from the modelization made previously on the diluted systems. In the high-Q domain both curves nicely superimpose, indicating that the estimated contrast values extracted from the fits and experimental silica volume fraction (determined by TGA) are consistent, and scale as $Q^{-4}$, showing the single silica spheres scattering. The most significant difference is the apparition, for both systems, of a maximum in the low-Q domain, which can be highlighted by representing the inter aggregate structure factor $S_{inter\ agg}(Q)$ (Figure 5(b)), obtained by dividing the total scattering intensity I(Q) by the form factor of the primary particles P(Q) deduced from the analysis of the diluted solution (Figure 1). This maximum can be commonly interpreted as the formation of a connected network presenting a regular quasi periodic mesh size [5, 6, 25]. The typical size of the network is deduced from the position of the maximum Q*, which corresponds in the real space at a distance of D~2π/Q* and gives respectively D~64 nm and D~100 nm for the un-grafted and for the grafted system. Thus, for the un-grafted system, the maximum corresponds to a smaller mesh size while the grafted one to a larger network mesh size. As for the diluted case, this difference can be explained by the local particle-particle interactions, illustrated by the strong correlation hole, which are stronger for the un-grafted system and lead to a tighter network. For the grafted system, these particle-particle attractions are reduced by the grafted polymer layer, leading to the formation of less contracted network.

[Figure 6]

The dispersion at larger scale is probed on the same samples (15.4% v/v for the un-grafted system in Figure 6(a) and 11.6% v/v for the grafted case in Figure 6(b)) with TEM images. Both pictures show a similar organization over a large range of scales and confirm the formation, for both un-grafted and grafted systems, of a connected filler network at high silica concentration. They also confirm that interactions appear more pronounced for the un-grafted case and are attenuated by the polymer layer for the grafted system. At larger scale, the apparent structure of the formed network seems indeed to be tighter for the un-grafted case than for the grafted one, with typical mesh sizes consistent with the values deduced from SANS.

*4. Mechanical measurements*

Mechanical properties of un-grafted and grafted silica nanocomposites were investigated both at low deformation (by plate/plate rheometer (Ares)) and large (uni-axial stretching) ones as a function of the silica volume fraction $\Phi_{SiO2}$.

[Figure 7]

First, at low deformation, Figure 7(a) presents the evolution of elastic modulus G' as a function of pulsation $\omega a_T$ for different grafted-silica volume fractions (4.3%, 8.1% and 11.6% v/v of silica). At high and intermediate $\omega a_T$, the curves superimpose with the reference one: no differences are observed whatever the silica content. In this regime (Rouse domain), short relaxation times of polymer matrix or small chains are observed. This behavior is not affected by the presence of fillers. At low $\omega a_T$ (i.e. long relaxation times associated with longer chains or particles relaxation), we observe a progressive increase of G' with $\Phi_{SiO2}$. The terminal time of nanocomposites is increased compared to the pure polymer one. For $\Phi_{SiO2}$=4.3% v/v, the terminal time is shifted at lower $\omega a_T$ and still observed in our experimental time range. As soon as the silica fraction reaches 8.1% v/v, the terminal relaxation time becomes so long that the creep zone is no longer visible in the experimental window. An additional elastic contribution is also observed and more and more pronounced when we increase silica volume fraction, finally tending at $\Phi_{SiO2}$=11.6%v/v to a plateau close to $10^5$ Pa.

These observations are close to the ones observed for the un-grafted system (already discussed in reference [25]) and presented in insert on Figure 7(a), except that for the grafted samples we can explore higher silica content (up to 11,6%v/v). Indeed, by this plate/plate technique for un-grafted systems, we only focused on the samples with low silica concentration (from 0 to 5% v/v), because for higher silica fractions (>5% v/v) the measurement reproducibility was not ensured (risk of slippage and too high modulus). From 3 to 5% v/v, the additional (with respect to pure matrix) elastic behavior is even more marked in the same low $\omega a_T$ range. The trend of a plateau at lower $\omega a_T$ is more and more apparent at $\Phi_{SiO2}$ = 4% and 5% v/v. This supports the idea that in grafted nanocomposites the reinforcement is less important at high $\Phi_{SiO2}$. The Figure 7 (b) presents the elastic and loss moduli (G' and G'') at one concentration where grafted and un-grafted composites can be directly compared (around 4% v/v in silica). All the curves superimpose perfectly, indicating no differences in behavior either for short or long relaxation times. The existence of a long elasticity plateau is accompanied, predictably, by a low dissipation, so that the G'' values become inferior to the G' values in the low frequency range and the slopes of G' and G'' curves become parallel as observed in gel process [6]. In this range of concentration the mechanical behavior is similar with or without grafted chains.

Now we investigate the reinforcement at higher deformation by uni-axial stretching: by this technique we can study samples filled with larger silica volume fractions. Figure 8 shows the evolution of stress as a function of elongation ratio λ for different silica volume fractions $\Phi_{SiO2}$, for the un-grafted (blue squares) and grafted (green curves circles) systems: (a) 4% (un-grafted) / 4.3% (grafted) (b) 7.5% (un-grafted) / 8.1% (grafted), (c) 12.5% (un-grafted) /11.6% (grafted). As expected, for both systems, we observe that the stress of the nanocomposites increases with silica volume fraction compared to the unfilled polymer (our reference, black triangles). At identical volume fraction several behaviors are observed.

**[Figure 8]**

At low $\Phi_{SiO2}$ (Figure 8(a), ~4%v/v) the curves superimpose well, the stress is higher than for the unfilled matrix but keeps the same shape and evolution on the whole λ scale. Grafting does not influence the mechanical properties. For higher concentrations around ~8% v/v on Figure 8(b), the stress for the filled samples is now significantly higher compared to the unfilled matrix with changes of curves as a function of the elongation: the initial increase of stress and the following break in slope are more pronounced for the un-grafted system than for the grafted one. The most significant effect appears for the highest concentration, ~12% v/v, presented in Figure 8(c), with a shoulder at low elongation clearly visible for the un-grafted while only a breaking slope is observed on the stress curve for the grafted system. The initial slope of stress at low λ is very different between the grafted and the un-grafted system, indicating that more energy is needed to deform the un-grafted system than the grafted one.

To emphasize this behavior, we divide the effective Young modulus (initial slope of stress/strain curves) of the nanocomposites by the one of the pure matrix. In figure 9, which plotting this reinforcement factor as a function of the silica volume fraction, we observe a strong increase at high $\Phi_{SiO2}$ illustrating at once the main mechanical difference between the two systems.

**[Figure 9]**

In this plot, we observe a first linear part identical whatever the system is (un-grafted or grafted), followed by a divergence of reinforcement more pronounced for the un-grafted systems. At high λ (Figure 8), the curves come back close to each other and follow the same stress evolution. They appear to be simply shifted compared to the reference curve by a constant value increasing with silica volume fraction. Many equations have been used to calculate the reinforcement of nanocomposites, like Guth and Gold equations [32] (black dash line in Figure 9), but these theoretical predictions do not take into account the direct contact between particles and therefore underestimate the reinforcement of both systems at large volume fractions.

Thus, to summarize, at low silica volume fraction the mechanical properties are identical for each systems, but they become different at high $\Phi_{SiO2}$ with a strong increase of elastic modulus G' at low $\omega a_T$ and a fast divergence of reinforcement factor for the un-grafted system. These results indicate that the interfacial interaction modifies the mechanical properties when the silica volume fraction is high enough. These behaviors will be discussed below from the perspective of fillers arrangement, polymer/filler and filler-filler interfacial interaction.

**Discussion**

Our results give a quantitative correlation between interfacial interaction and mechanical reinforcement by the study of model systems in which a direct comparison can be performed between PS-grafted silica nanocomposites and their un-grafted equivalents. For the grafted system, we firstly characterized the grafted particles in solution with SANS and contrast variation method to check the efficiency of our grafting technique and the preservation of dispersion stability. The direct comparison of nanocomposites filled with grafted or un-grafted nanoparticles is not obvious, because the local organization of the particles inside the polymer matrix will depend mainly on the grafting characteristics (grafting density and grafted-to-free chains length ratio) [15, 16, 17, 18, 33]. The grafted layer can modify the aggregation processes and lead to a final dispersion different from the one obtained with un-grafted particles: the distinction between the filler contribution and the polymer-filler interfacial contribution on the mechanical properties of the materials remains thus greatly limited. However, contrary to such expected behaviors, the results reported here provide a first surprising conclusion: the dispersion of the grafted and the un-grafted silica particles inside the PS matrix are very similar: two main regimes are obtained as a function of the silica volume fraction $\Phi_{SiO2}$. A dilute one, (below $\Phi_{SiO2}$ = 5% v/v) in which the silica particles arrange in non-connected finite size aggregates, and a concentrated one (above $\Phi_{SiO2}$ = 5% v/v) in which the aggregates form a continuous connected network arising from a kind of percolation. Therefore here the grafted layer does not influence significantly the aggregation process compared to the un-grafted case. Beyond the influence of the grafted-to-free chains interactions on the particle dispersion, it is now admitted that sample processing method can also affect the final fillers dispersion [34]. We thus guess here the similarity in the dispersion obtained for both systems is the result of the use of a strictly identical sample processing method. This influences the aggregation process more than the grafting.

The structure factor S(Q) shows that the particle-particle interaction is strong for the un-grafted systems but remains weaker for the grafted one. According to recent results [35, 36], the particle-particle interaction potential is rather soft for grafted systems compared to the classical hard sphere models. As a result, even with a comparable size ($N_{agg}$ = 9 for un-grafted and $N_{agg}$ =7 for grafted), in the diluted regime, the un-grafted aggregates appear more compact ($D_f$ =1.8) than the grafted aggregates ($D_f$=1.1). At the same time, in the concentrated regime, the filler network appears to be tight for the un-grafted case and rather less contracted for the grafted case. Thus the two nanocomposites systems behave similarly with silica dispersion, the dispersion state remains unchanged by polymer grafting. The identity in filler structure opens the way to a direct comparison of the mechanical properties investigated over a large range of deformation as a function of the silica volume fraction $\Phi_{SiO2}$.

***At low silica volume fraction (small non-connected aggregates)***, both systems undergo the same mechanical behavior: both show a solid-like behavior (long relaxation time) in the linear regime. Mechanical reinforcement is identical whatever the system and thus does not seem to depend on interfacial interactions. This behavior is confirmed by uni-axial stretching measurements at 5%v/v (see Fig. 8(a)).

One common explanation is that such solid-like behavior is the result of percolation of filler through different processes involving directly the polymer-filler interface [21]. For the un-grafted system, this process would be provided by a "glassy layer" surrounding the filler [6, 21, 37], while for the grafted system it is given by contact between grafted layers [16, 17, 19, 38]. These grafted layers could also be glassy, as mobility of grafted chains in the interface could be reduced [18]. In this view, the morphology of the aggregates (a dispersion of individual particles does not give comparable solid-like effect amplitudes [16, 17, 18]) is also supposed to play a role especially in the quantity of polymer chains involved [39]. However, our results show that a strong modification of the polymer-filler interface with grafting (associated with a small modification of the aggregates morphology) does not influence the solid-like response of the materials. This observation suggests that typical scales larger than the one of the polymer-filler interface are involved. This idea is supported by our observations in both systems in which we can see that the characteristic distance between the aggregates (~100 nm) is larger than the typical size of the polymer-filler interface (~10 nm). Such longer range process can include different mechanisms like chain dynamic effect, "glassy" path [40] or filler self relaxations [37] which remains to be investigated with complementary specific experiments. Roberston and Rackaitis [41] have recently proposed that such mechanical behavior could be interpreted in term of suppression of flow relaxation (chain diffusion/reputation) of polymer chain due to long range interaction with the particles.

***At high silica volume fraction (connected network)***, we observe, for both grafted and un-grafted systems, the appearance of a fast increase of the reinforcement factor (a deviation from hydrodynamic behavior) above a given silica volume fraction (Figure 9), which suggests the formation of a percolating-like filler network. This observation is in good agreement with the filler structure determined by TEM (Figure 6) and SANS (maximum of the intensity in the low-Q domain in Figure 5). Such behavior is rather classical for fractal fillers and has already been observed in many systems [5, 25, 42, 43, 44]. The original point here is the strength of reinforcement, significantly larger for the un-grafted system than for the grafted one. We can relate this observation to the local structure of the two networks: tight for the un-grafted and less contracted for the grafted one. Contrary the previous case, above the percolation threshold, the mechanical behavior depends then on the interfacial polymer-filler interactions and can be directly correlated with the particle-particle interactions strength: the stronger the interfacial interaction (as seen without grafting from the correlation hole in the S(Q) in Figure 4(b)), the higher the reinforcement. For the grafted case, reinforcement is lower because the strength of the filler-filler interactions is lower, due to the layer of grafted polymer which yields a softer contact between particles. The grafted sample is homogeneously deformable on a larger deformation range. This is particularly visible on the stress-strain isotherm (Figure 8 (c)), where we observe the maximum of stress at low λ for the un-grafted system while only a change of slope is seen with grafting at similar volume fraction. This strong distortion of the filler network seems to be avoided owing to the deformability of the grafted chains. This "soft layer" behavior appears to be in opposition with glassy layer interfacial effects [45], and could be more investigated with further direct determination of the grafted layer deformation associated with measurements of the grafted chains dynamics.

**Summary and Conclusion**

In this paper we directly investigated the effect of interfacial interactions on mechanical reinforcement by using two models PS/silica nanocomposites differing only by the state of the silica particles surfaces, raw or grafted with PS chains, but close from the dispersion point of view. To establish the latter, rather important point, the small scale structure of the nanocomposites is investigated using a combination of SANS measurements and TEM imaging, while the mechanical behavior of the materials is probed over a wide range of deformations as a function of the silica volume fraction. We observe two regimes separated by a connectivity threshold and different range of particle-particle interaction. At low silica volume fractions, both systems arrange in non visibly connected fractal aggregates and present a similar solid-like behavior illustrating that reinforcement is independent here of the local polymer-particle interfacial interaction and suggesting mechanisms at sizes larger than the chain one. Conversely, at high silica volume fraction, the aggregates form a continuous network inducing a fast increase of the reinforcement factor whose amplitude is then clearly system-dependent. Reinforcement can now be directly related to the local state of the interface and is higher for strong particle-particle interaction and weaker when the interaction is modulated by the grafted polymer layer. Lower reinforcement and enhanced deformability for grafted nanocomposites imply local deformation of the polymer chains grafted at the surface of the particles. The local interfacial polymer-filler interaction, tunable with grafting, appears then to be a relevant parameter to control the strength of the reinforcement at large filler volume fraction, and an irrelevant one at low volume fraction when fillers are not directly connected. This result opens the way to improvements of both fundamental models and applicative processes by driving the choice of the computational length scale parameters (short or long range filler interactions) and the one of the nature of the composite components as a function of the expected macroscopic mechanical properties.

**ACKNOWLEDGMENTS**

The authors thank Dr. Didier Gigmes and Prof. Denis Bertin (CROPS, Marseilles, France) for their contribution to the grafting synthesis and Dr. Sylvère Said (LIMATB, Lorient, France) for useful discussions on mechanical properties. Commissariat à l'Energie Atomique (CEA) is acknowledged for PhD Grant of C. Chevigny and we thank the Région Bretagne and the CEA for the support of N. Jouault's PhD Grant. The authors thank Florian Meneau for SAXS beam time on Swing at French Synchrotron Soleil.

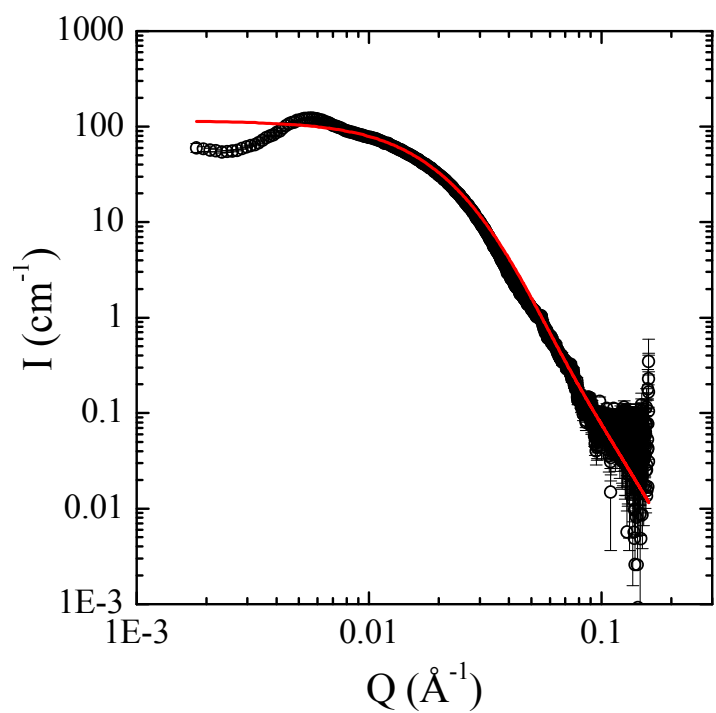

Figure 1: SAXS curve of a diluted solution (0.1% v/v) of native silica particles dispersed in DMAc. The full red line is the result of the fit with a polydisperse spheres form factor with a mean radius R = 6 nm and a polydispersity $\sigma_{\text{log-normal}} = 0.3$.

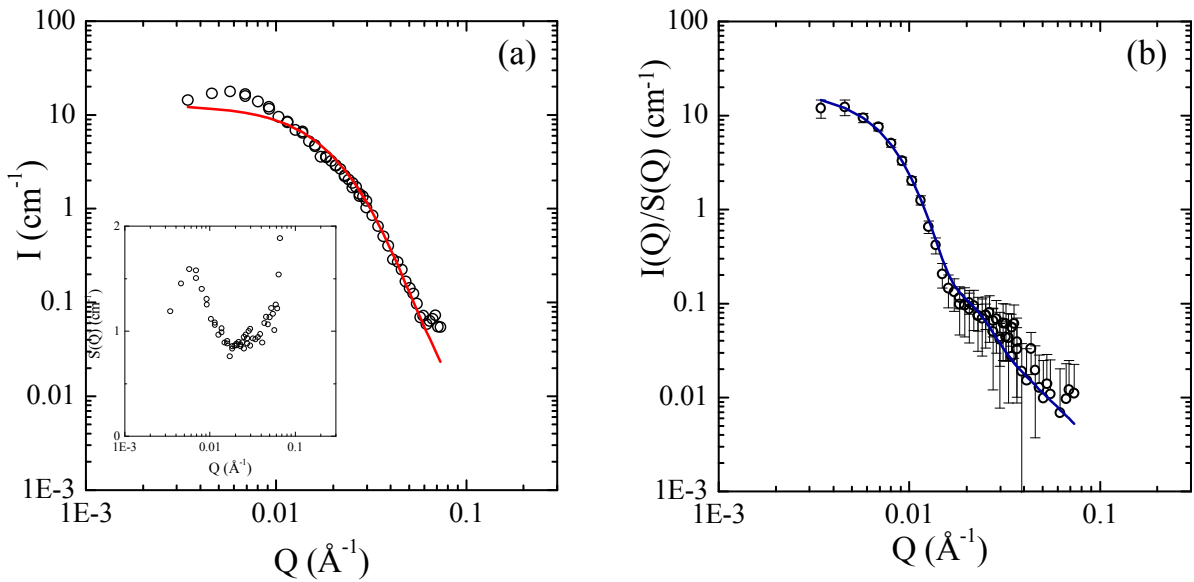

Figure 2: (a) SANS curve of PS-grafted silica nanoparticles in solution in polymer contrast-matching conditions (open circles), the full red line is the calculated form factor of the native silica particles and in insert the structure factor S(Q) deduced by dividing the total intensity by the silica form factor. (b) SANS curve of PS-grafted silica nanoparticles in solution in silica contrast-matching condition (open circles), the full blue line is the best fit obtained with the Gaussian chain model corresponding to the grafted chains parameters (see text for details).

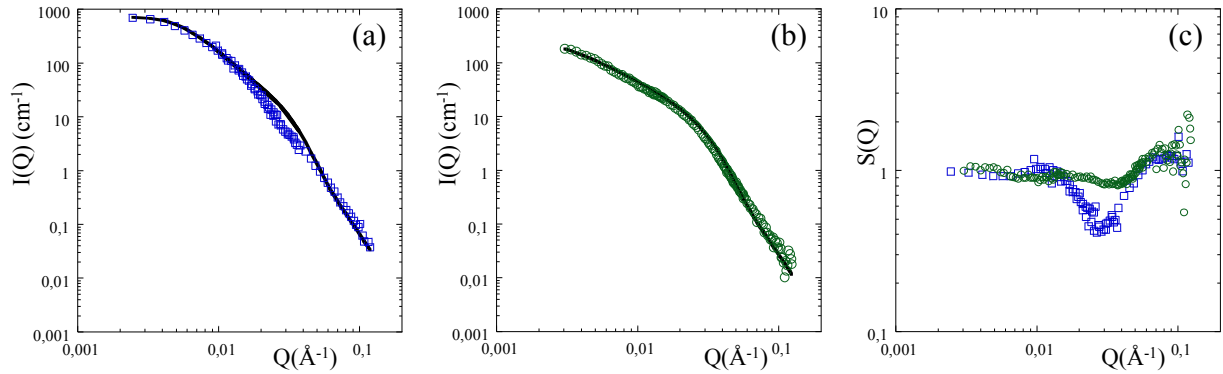

Figure 3 : (a) SANS of nanocomposite filled with un-grafted silica particles at a silica volume fraction of 6.65% v/v (blue squares), the full black line is the best fit using the model of fractal aggregate which gives $N_{agg}$ = 9 and $D_f$ = 1.8. (b) SANS of nanocomposite filled with grafted silica particles at a silica volume fraction of 4.30% v/v (green circles), the full black line is the best fit using the model of fractal aggregate which give $N_{agg}$ =7 and $D_f$ = 1.1. (c) Structure factor S(Q) of un-grafted (blue square) and grafted (green circles) systems obtained by dividing the scattered intensity by the form factor of the aggregates.

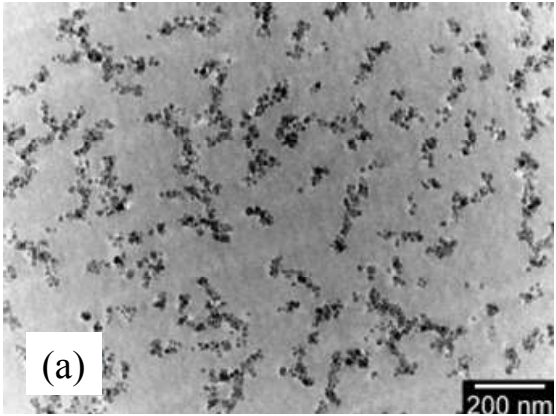 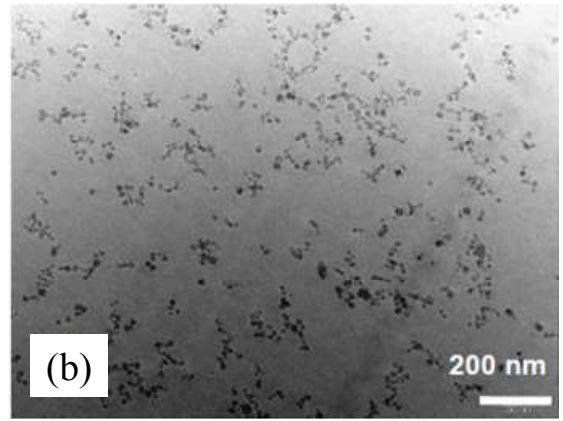

Figure 4: (a) TEM picture of the un-grafted nanocomposite filled with 6.65 % v/v of silica particles (black is silica and grey is polymer). (b) TEM picture of the grafted nanocomposite filled with 4.30 % v/v of silica particles.

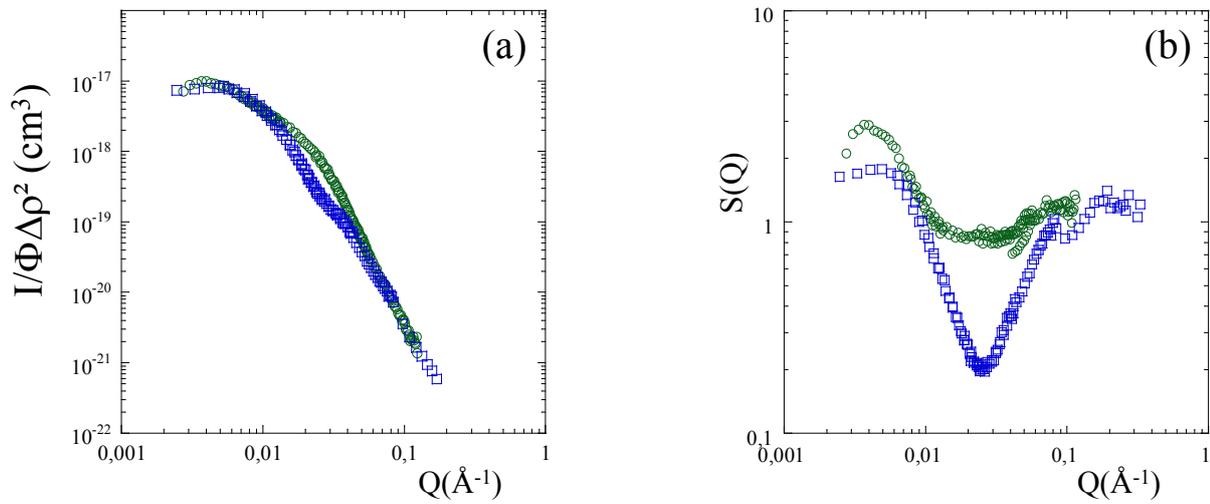

Figure 5: (a) SANS curves for un-grafted (blue square) at 15.40 % v/v of silica particles and for grafted (green circle) at 11.60 % v/v of silica particles normalized by the contrast term deduced from the modelization of the aggregates form factor, respectively equal to $\Delta\rho^2 = 3.88\ 10^{20}\ cm^{-4}$ and $3.00\ 10^{20}\ cm^{-4}$. (b) Structure factor S(Q) for un-grafted (blue square) and grafted (green circle) systems obtained by dividing the scattered intensity by the aggregate form factor.

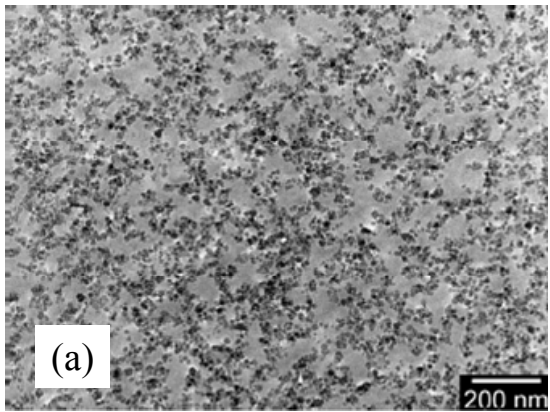 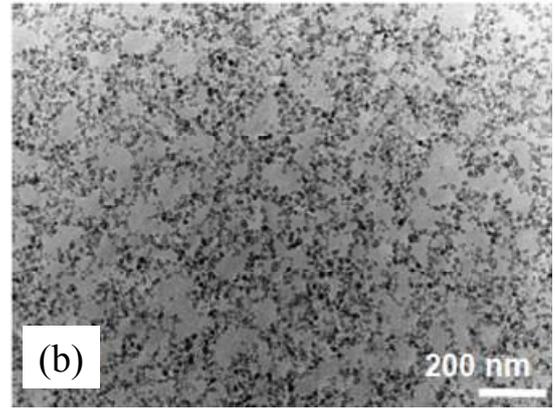

Figure 6: (a) TEM picture of the un-grafted nanocomposite filled with 15.40 % v/v of silica particles (black is silica and grey is polymer). (b) TEM picture of the grafted nanocomposite filled with 11.60 % v/v of silica particles.

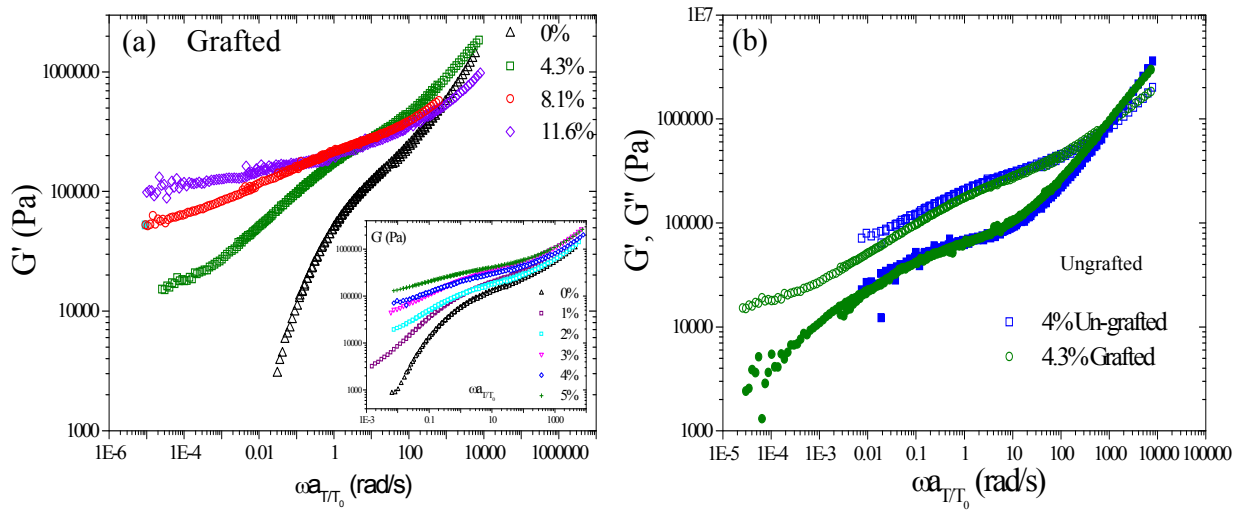

Figure 7: (a) Elastic modulus G' for the whole scale of concentrations (0, 4.30, 8.10, and 11.60 %v/v in silica) for the grafted case. G' for un-grafted case is given in insert (from 0 to 5%v/v in silica). (b) Comparison of the elastic and loss modulus (G', hollow marks, and G'', full marks) for the lowest silica concentration (4%, un-grafted, 4.3%, grafted).

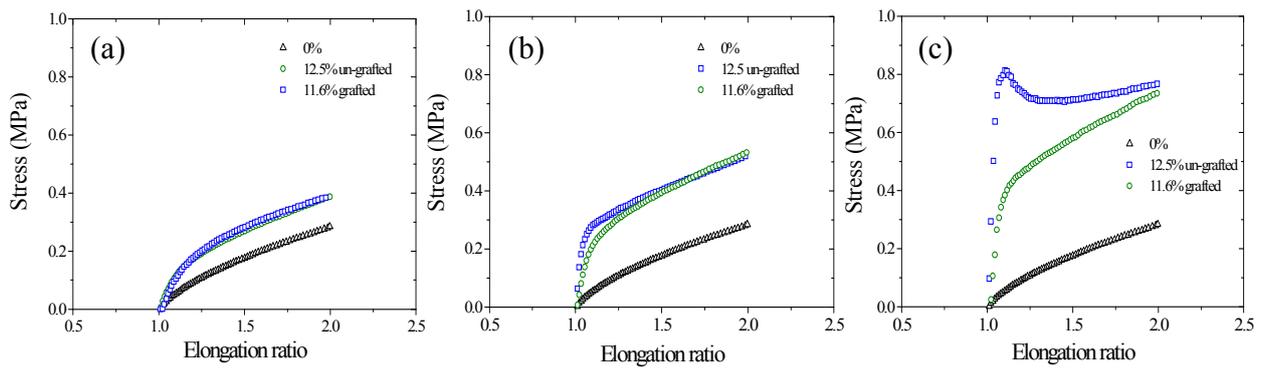

Figure 8: Stress as a function of elongation ratio: direct comparison between grafted (◯) and un-grafted (☐) cases: from the lowest silica volume fractions ((a), 4.00% un-grafted 4.30% grafted) to higher volume fractions ((b), 7.50% un-grafted, 8.10% grafted, and (c), 12.50% un-grafted, 11.60% grafted). Each time the stress curve for the pure polymer is also plotted (△).

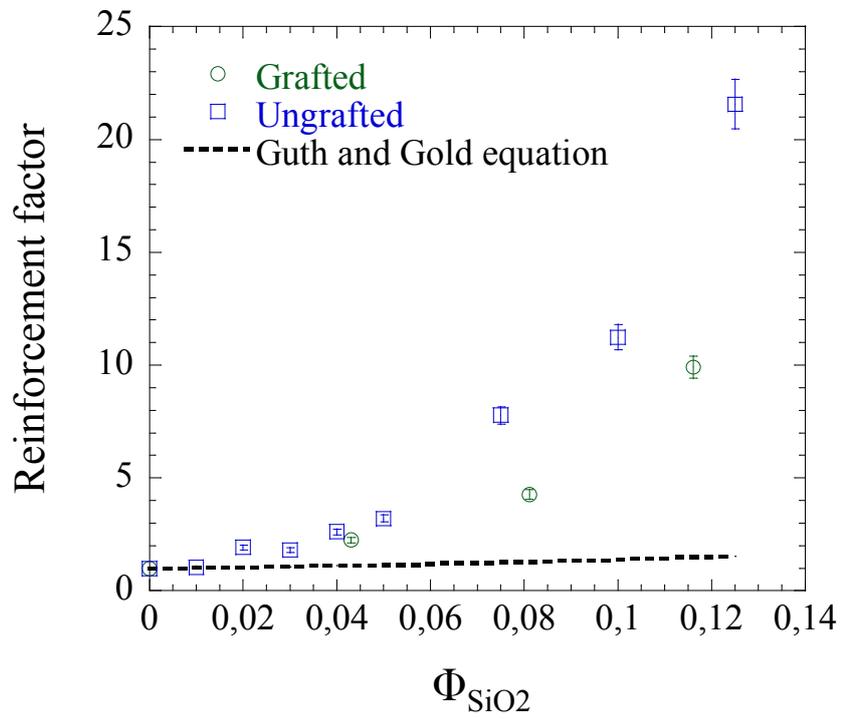

Figure 9: Relative reinforcement from stress/strain curves as a function of silica volume fraction for grafted (○) and un-grafted (□) silica nanocomposites. The black dash line corresponds to the Guth and Gold predictions for spherical fillers.

SYNOPSIS TOC

The effect of polymer/filler interfacial interaction on mechanical properties has been investigated on un-grafted and PS-grafted silica nanocomposites presenting similar filler dispersion. Below filler connectivity, reinforcement is similar for both systems and thus independent of polymer-filler interface state. When direct silica connectivity occurs, the grafted polymer layer yields a soft network between particles decreasing the strength of the connectivity. Our results suggest that the reinforcement is clearly modulated by the strength of the filler/filler interaction.

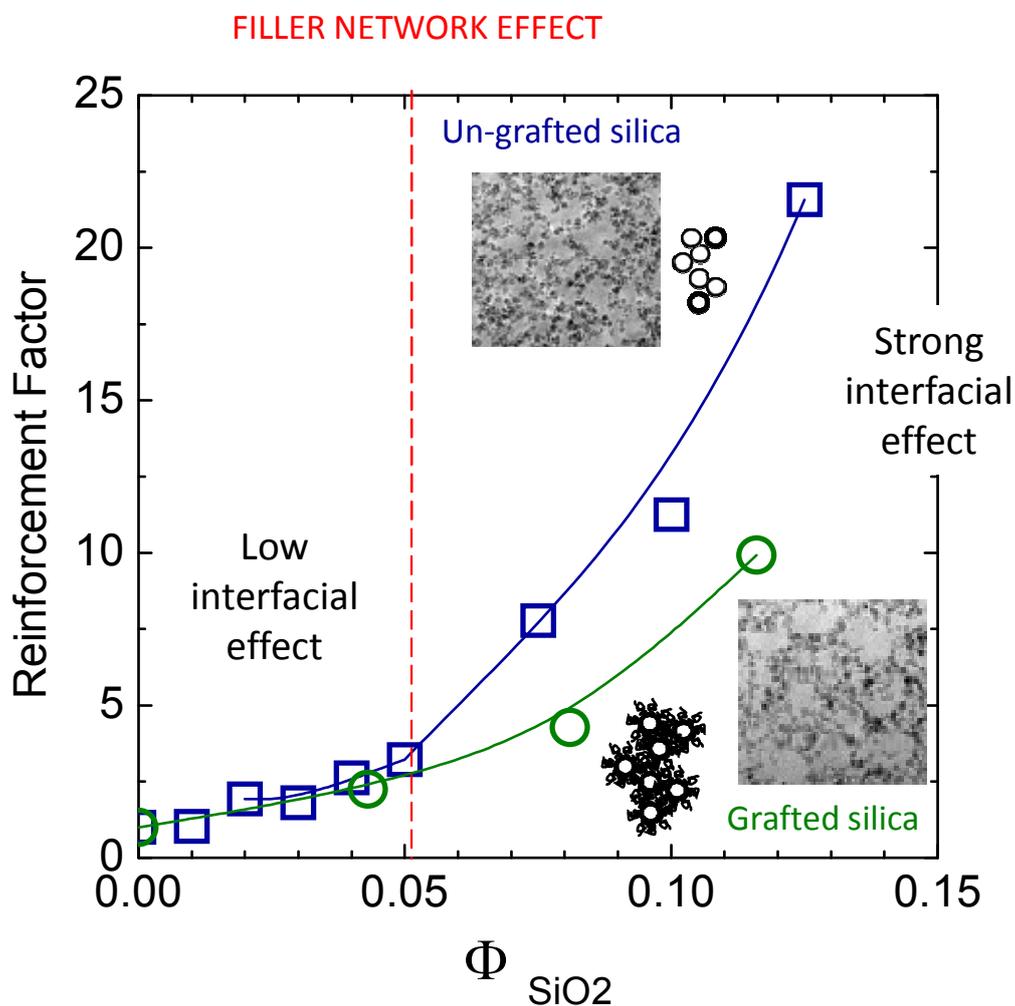